%% file: icassp_template.tex
\documentclass[9pt]{extarticle}
\usepackage{extsizes}
\usepackage{spconf,amsmath,graphicx}

\usepackage{hyperref}
\usepackage{url}

\usepackage{booktabs, multicol, multirow,array} %
\newcolumntype{?}{!{\vrule width 1pt}}
\newcolumntype{|}{!{\vrule width 0.4pt}}
\usepackage{arydshln}

\usepackage{graphicx}
\usepackage{amsmath}
\newcommand\norm[1]{\left\lVert#1\right\rVert}
\usepackage{multirow,stfloats}
\input{math_commands.tex}

\usepackage{xcolor}

\let\oldbibliography\thebibliography
\renewcommand{\thebibliography}[1]{%
  \oldbibliography{#1}%
  \setlength{\itemsep}{0.0em}%
}

\usepackage{etoolbox}
\newtoggle{prepub}
\toggletrue{prepub}

\iftoggle{prepub}{
    \usepackage{fancyhdr, textcomp}
    \fancypagestyle{IEEE_Copyright_footer}{%
        \fancyhf{}
        
        \fancyfoot[L]{\fontsize{9}{11} \selectfont {\textcopyright} 2021 IEEE. Personal use of this material is permitted. Permission from IEEE must be obtained for all other uses, in any current or future media, including reprinting/republishing this material for advertising or promotional purposes, creating new collective works, for resale or redistribution to servers or lists, or reuse of any copyrighted component of this work in other works.}
    }
    \pagestyle{empty}
}

\title{Multi-Channel Speech Enhancement using Graph Neural Networks}
\name{Panagiotis Tzirakis,
             Anurag Kumar and
             Jacob Donley}
\address{Facebook Reality Labs Research, USA \\
         {panagiotis.tzirakis12@imperial.ac.uk}, \{anuragkr, jdonley\}@fb.com}

\begin{document}
%
\maketitle
\iftoggle{prepub}{\thispagestyle{IEEE_Copyright_footer}}{\thispagestyle{empty}}
\begin{abstract}
Multi-channel speech enhancement aims to extract clean speech from a noisy mixture using signals captured from multiple microphones. Recently proposed methods tackle this problem by incorporating deep neural network models with spatial filtering techniques such as the minimum variance distortionless response (MVDR) beamformer. In this paper, we introduce a different research direction by viewing each audio channel as a node lying in a non-Euclidean space and, specifically, a graph. This formulation allows us to apply graph neural networks (GNN) to find spatial correlations among the different channels (nodes). We utilize graph convolution networks (GCN) by incorporating them in the embedding space of a U-Net architecture. We use LibriSpeech dataset and simulate room acoustics data to extensively experiment with our approach using different array types, and number of microphones. Results indicate the superiority of our approach when compared to prior state-of-the-art method.
\end{abstract}

\begin{keywords}
Speech enhancement, deep learning, multi-channel processing, graph neural networks
\end{keywords}

\section{Introduction}
\vspace{-0.1in}
\label{introduction}

Humans can naturally focus their auditory system to attend on a single sound source while cognitively ignoring other sounds. The exact mechanism that the brain employs to perform such task in difficult noisy scenarios, often termed the \textit{cocktail party problem}~\cite{cherry1953some}, is still not completely understood. However, studies have shown that binaural processing can help alleviate this problem~\cite{hawley2004benefit}. Spatial information helps the auditory system group sounds from specific directions and segregate them from other directional interfering sounds.

Multi-channel speech enhancement is the process of enhancing a target's speech corrupted by background interference using multiple microphones. It is very crucial to many applications including, but not limited to, human-machine interfaces~\cite{bai2013acoustic}, mobile communication~\cite{tan2019real}, and hearing aid~\cite{nossier2019enhanced, nugraha2016multichannel}. 

While the problem has been studied for a long time, it remains a challenging one. The target's speech signal can be corrupted by not only other sound sources but also by reverberation from surface reflections. Traditional approaches include spatial filtering methods~\cite{benesty2008microphone, gannot_consolidated_2017} that often make use of the spatial information from the sound scene, such as angular position of the target's speech and the microphone array configuration. These approaches are regularly termed \emph{beamforming}, a linear processing model that weights (“masks”)  different microphone channels in the time-frequency domain in order to suppress source signal components that are not the target sound source. In the case of the minimum variance distortion-less response (MVDR)~\cite{souden2009optimal} beamformer, first the desired source transfer function and noise covariance matrices are estimated, often via power spectral density (PSD) matrices, then beamforming weights are computed and applied to the signals. Although these approaches can perform well, their performance depends on reliable estimation of spatial information, which can be challenging to accurately estimate in noisy conditions. 

Deep neural networks (DNN) have been widely used in variety of audio  tasks, such as emotion recognition~\cite{tzirakis2017end}, automatic speech recognition~\cite{besacier2014automatic}, and speech enhancement and separation ~\cite{wang2018supervised}. For multi-channel processing, they have been incorporated with traditional spatial filtering methods, such as the conventional filter-and-sum beamformer. This has mainly been accomplished in two ways, both of which are applied in the frequency domain. In one approach, a DNN is used to predict directly the beamforming weights~\cite{sainath2017multichannel}. In a second approach, a DNN is used to estimate a mask which is applied to the short-time Fourier transform (STFT) of the signal such that the PSD matrices are computed~\cite{chakrabarty2019time}. Then, a beamforming method is applied, such as MVDR, that computes the filter coefficients using the PSD matrices~\cite{higuchi2018frame, wang2018all, chakrabarty2019time}. These methods use DNN's in different ways, however, the end goal for each of them is the same, which is to predict the filter coefficients. Recently, however, a shift in the audio community has emerged towards incorporating attention mechanisms in the deep neural network architectures~\cite{tolooshams2020channel} to implicitly perform spatial filtering.

In this paper, rather than using traditional beamforming method with a DNN or attention mechanism, we propose a novel approach for multi-channel speech enhancement and de-reverberation. In particular, we view each audio channel as lying in a non-Euclidean space, more specifically, a \textit{graph}, which is learned from the observations. Formulating the problem in such a manner allows us to exploit methods from the graph neural network's (GNN) domain~\cite{bronstein2017geometric}, and perform our training in an \textit{end-to-end} manner. In addition, learning a graph structure allows the network to adapt its structure according to the dynamic sound scene. To the best of our knowledge, this is the first method which formulates multi-channel speech enhancement and de-reverberation through a graph and uses graph neural networks to solve it. 

Our approach relies on both real and imaginary parts of the complex mixture in the short-time Fourier transform (STFT) domain and estimates a complex ratio mask (CRM) for a reference microphone. The CRM is then applied to the mixture STFT to obtain the clean speech. We apply our proposed method for simultaneous speech enhancement and de-reverberation tasks. To this end, we simulate data leveraging the LibriSpeech~\cite{librispeech} dataset. In particular, we simulate data for different microphone array configurations - \emph{linear, circular, and distributed} while varying the number of microphones. We use Short-Time Objective Intelligibility (STOI)~\cite{taal2011algorithm}, Perceptual Evaluation of Speech Quality (PESQ)~\cite{rix2001perceptual} and Signal to Distortion Ratio (SDR)~\cite{vincent2006performance} as evaluation metrics in our experiments. We also show that our approach outperforms a recently proposed neural network based multi-channel speech enhancement method.

\begin{figure*}[!t]
\centering
    \includegraphics[width=0.75\linewidth]{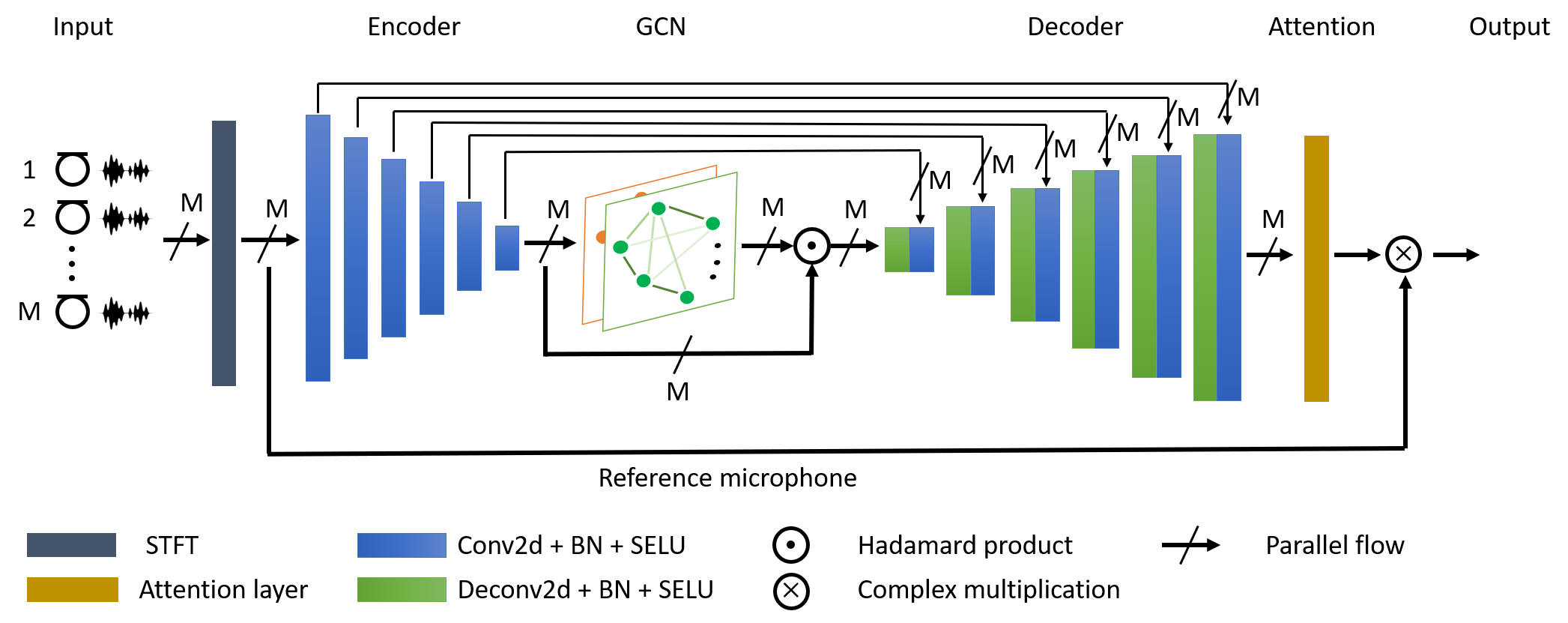} 
    \caption{The proposed model to multi-channel speech enhancement using graph neural networks. The complex spectrogram of each microphone signal is computed and passed to the encoder. The extracted representations of each channel are passed to a graph convolution network for spatial feature learning. The extracted features of each channel are passed to a decoder, and a weighted sum of the decoder output is performed. The output is (complex) multiplied with a reference microphone complex spectrogram to produce the clean spectrogram.}
    \label{fig:approach}
\end{figure*}

\section{Graph neural networks}
\label{gnn}

Graph neural networks (GNN)~\cite{bronstein2017geometric} are a generalization of conventional neural networks, designed to operate on non-Euclidean data in forms of graph. Graphs provide considerable flexibility in how the data can be represented and
structured and GNNs allow one to operate and generalize neural network methods to graph structured data .  A particular type of GNN is the convolutional GNNs which is based on the principles of learning through shared-weights, similar to convolutional neural networks (CNNs)~\cite{kipf2016semi}.  Broadly, there are two approaches for building graph convolutional neural networks, \emph{Spectral} GCNs and \emph{Spatial} GCNs~\cite{wu2020comprehensive, velivckovic2017graph,balcilar2020bridging}. Spectral GCNs are based on principles of spectral graph theory. More specifically, the graph processing is based on Eigen-decomposition of graph Laplacian, which are used to compute the Fourier transform of a graph signal, through which graph filtering operations are defined. Spatial GCNs define convolutions directly on graph data and tries to capture information by aggregating information from neighboring nodes through shared weights. Spatial GCNs are computationally less complex as well as generalize better to different graphs. While Spectral GCNs operate on fixed graph Spatial GCNs have the flexibility of working locally on each node without taking into account the full fixed graph. They do however require node ordering. 

A key aspect of our proposed method is that we construct the graph dynamically conditioned on the task at hand. This dynamic graph construction approach enables the framework to capture multi-channel information for each audio in a sample specific manner.


\section{Multi-channel graph processing}
\label{gap}

Our proposed framework is schematically illustrated in Fig.~\ref{fig:approach}. The audio signal from each channel is transformed to a time-frequency (T-F) representation which are fed to the neural network framework. The inputs are first passed through an encoder network which tries to learn higher level features from the inputs~(Sec~\ref{representation_learning}). These feature representations for the audio channels are then used to construct a graph which captures the multi-channel information through its nodes and edges. At this point, we use GCNs~(Sec~\ref{gcn}) to aggregate information from each microphone. The output representation of each node in the graph is passed as input to the decoder, which then transform the signals back to their original dimensions. Finally, a weighted sum of the decoder outputs is computed, and it is (complex) multiplied with the STFT of a reference microphone to compute the clean STFT.

\subsection{Audio Representations Learning}
\label{representation_learning}
The first major step in our proposed framework is to learn representations for audio signals from each microphone. To do this, the audio signals are first converted to a time-frequency representation through Short-Time Fourier transform (STFT). The real and imaginary parts of this complex are stacked together to obtain a 2 channel tensor of size $2 \times T \times F$, where where $T$ represents the total number of time segments, and $F$ represents the total number of frequency bins. Considering all $M$ channels, it leads to a $M \times 2 \times T \times F$ dimensional input to the framework.

We utilize a U-Net architecture ~\cite{badrinarayanan2017segnet} to learn representations for the inputs. U-Net based architectures has been shown to work well for speech enhancement~\cite{tanconvolutional}. Complex spectrograms from each channel are passed to the encoder. The representations produced by from the encoder are used to obtain the multi-channel graph. The decoder outputs $M$ tensors of the same dimension as the input. We combine these tensors in a unified representation using attention layer, i.\,e., a weighted sum of them.


\subsection{Graph Construction}
\label{graph_construction}
Traditionally, signal processing based methods have been used to extract information from audio signals captured by multiple microphone. Here, we propose a \emph{novel} approach which extracts multi-channel information through graph processing. The first step in this process is to construct a graph using the audio representations for different channels obtained in the previous step.  

We construct an undirected graph, $\mathcal{G}=(\mathcal{V}, \mathcal{E})$, where $\mathcal{V}$ represents the set of nodes, $v_i$ (i.\,e. microphones), of the graph. $\mathcal{E}$ represents the edges, $(v_i,v_j)$, of the graph between two nodes. In graph theory, the graph is characterized by an adjacency matrix $\mathcal{A} \in \mathbb{R}^{|\mathcal{V}| \times |\mathcal{V}|}$, where $|\cdot|$ indicates the cardinality and a degree matrix $\mathcal{D}$ \cite{zhu2005semi}. We consider \emph{weighted} adjacency matrix, where the entries of $\mathcal{A}$ correspond to a weighted edge $(v_i,v_j) \in \mathcal{E}$ between two nodes of the graph. Intuitively, each weight represents a similarity between the feature vectors of two nodes in the graph. In our approach here, these weights, $w_{ij},\,\,\{i,j \in |V|\}$, 
of the adjacency matrix $\mathcal{A}$ are learned during the training process. For two nodes $v_i$ and $v_j$, we first concatenate their representations, $\mathbf{f_{v_i}}, \mathbf{f_{v_j}} \in \mathbb{R}^N$ as $[\mathbf{f_{v_i}}||\mathbf{f_{v_i}}]$ and then pass them through a non-linear function $F([\mathbf{f_{v_i}}||\mathbf{f_{v_j}}])$. We construct our adjacency matrix by normalizing the weights of each node to sum to one. The node degree matrix $\mathcal{D}$ is a diagonal matrix $\mathcal{D}_{ii} = \sum_{j}\mathcal{A}_{ij}$.

\subsection{Graph Convolution Network}
\label{gcn}
The graph, $\mathcal{G}$ constructed in the previous section, provides a structured way to capture the information from all microphones.  We can now exploit GCNs to learn spatial relations from this graph. We apply GCN to learn higher abstraction levels for the node features by learning representations for each node with respect to its neighbors. Given a graph $\mathcal{G}=(\mathcal{V}, \mathcal{E})$, the GCN applies non-linear transformation on the input feature matrix $\mathcal{X} \in \mathbb{R}^{|\mathcal{V}| \times N}$ with $N$ features. Mathematically, GCN can be represented as follows 

\begin{equation}
    \mathcal{H}^{(l)} =  g(\mathcal{D}^{-1/2} \mathcal{A} \mathcal{D}^{-1/2} \mathcal{H}^{(l-1)} \mathcal{W}^{(l-1)}).
\end{equation} 
\noindent
$\mathcal{H}^{(l)} \in \mathbb{R}^{|\mathcal{V}| \times K}$ is the $l^{th}$ layer with $K$ features with $\mathcal{H}^{(0)} = \mathcal{X}$. $\mathcal{D}$ is a diagonal node degree matrix, $\mathcal{W}^{(l-1)}$ is the trainable weight matrix at the $l-1^{th}$ layer, and $g$ is an activation function. 

\subsection{Loss Functions}
To train the overall framework, we consider loss computation in different forms. We consider loss computation through magnitude spectrogram, complex spectrogram and in raw signal domain. More specifically, following four losses are considered, \begin{equation}
    \mathcal{L}_{Mag} = \norm{\hat M - M}_1, \mathcal{L}_{spec} = \norm{\hat S - S}_1
\end{equation}
\begin{equation}
    \mathcal{L}_{Mag+Spec} = \mathcal{L}_{Mag} + \mathcal{L}_{spec} 
\end{equation}
\begin{equation}
    \mathcal{L}_{Mag+raw} = \mathcal{L}_{Mag} + \norm{\hat s - s}_1
\end{equation}
$\norm{\cdot}_1$ indicates the L1 norm, $M$, $S$ and $s$ indicate the magnitude spectrogram, complex spectrogram and clean signal, respectively, and $\hat{}$ indicates the corresponding predicted entities.

\begin{table*} [!t]
    \setlength{\tabcolsep}{4pt}\centering
    \begin{tabular}{lc?ccc?ccc?ccc}\toprule\midrule
         \multirow{2}{*}{Method} & \multirow{2}{*}{\# Mics} & \multicolumn{3}{c?}{Circular} & \multicolumn{3}{c?}{Linear} & \multicolumn{3}{c}{Distributed}\\
         \cmidrule(l{0.5em}){3-11}
          ~ & ~ & STOI & PESQ & SDR & STOI & PESQ & SDR & STOI & PESQ & SDR \\
          \midrule\midrule 
        \multirow{2}{*}{Noisy}
        & 2 & $0.67$ & $1.73$ & $0.51$  & $0.68$ & $1.75$ & $0.23$  & $0.66$ & $1.70$ & $0.35$ \\
        & 4 & $0.66$ & $1.62$ & $0.22$  & $0.65$ & $1.69$ & $0.59$  & $0.66$ & $1.64$ & $0.24$ \\
        \midrule
        \multirow{2}{*}{CRNN-C~\cite{chakrabarty2019time}}
        & 2 & $0.68$ & $1.75$ & $2.29$  & $0.68$ & $1.72$ & $2.17$  & $0.67$ & $1.73$ & $2.33$ \\
        & 4 & $0.70$ & $1.88$ & $4.89$  & $0.68$ & $1.75$ & $3.17$  & $0.67$ & $1.71$ & $2.25$ \\
        \midrule[1pt]
        \multirow{1}{*}{Proposed Single-Channel}
        & -- & $0.69$ & $1.98$ & $6.38$  & $0.69$ & $1.96$ & $6.48$  & $0.68$ & $1.94$ & $5.77$ \\
        \midrule
        \multirow{2}{*}{Proposed} 
        & 2 & $0.72$ & $2.20$ & $6.40$  & $0.72$ & $2.10$ & $7.03$  & $0.71$ & $2.04$ & $6.40$ \\
        & 4 & $0.73$ & $2.21$ & $8.53$  & $0.71$ & $2.10$ & $7.04$  & $0.71$ & $2.02$ & $6.72$ \\
    \end{tabular}
    \caption{Results (STOI, PESQ, SDR) of the enhanced signal for three array types, namely, circular, linear, and distributed configurations.}
    \label{sim_results}
\end{table*}

\section{Experiments}
\label{experiments}

\subsection{Dataset}

We utilize LibriSpeech dataset, which is a corpus of approximately $1000$\,h of English speech, captured in an anechoic chamber, with $16$\,kHz sampling rate. For our purposes, we use the dataset's sentences to simulate room acoustic data of three array types: linear, circular, and distributed. With the distributed array, we randomly place the microphones in the room. In addition, we experiment with $M\in\{2,4\}$ microphones in the array. The simulated observations consist of one speech signal mixed with $M-1$ noise signals, selected randomly from AudioSet~\cite{gemmeke2017audio} and located randomly in the room. The SNR levels of the mixed signals are from the set $\{-7.5, -5, 0, 5, 7.5\}$\,dB. The training set is comprised of 3 rooms with dimensions ($\mathrm{width}\times\mathrm{depth}\times\mathrm{height}$) from the set $\{3\times3\times2,~5\times4\times6,~8\times9\times10\}$\,meters, the development set of 2 rooms with dimensions from $\{5\times8\times3,~4\times7\times8\}$\,meters, and the test set of 2 rooms with dimensions from $\{4\times5\times3,~6\times8\times5\}$\,meters. All rooms have $RT_{60} = 0.5$\,sec. The training set is comprised of $6$\,h, and the development and test sets of $5$\,h.

\subsection{Experimental Setup}

We used the Adam optimization algorithm~\cite{kingma2014adam} to train the models  with a fixed learning rate of $10^{-5}$, and a mini-batch of size $20$ frames. The input representation is the complex STFT computed with a Hanning window of length $1024$, and frequency bins $513$, with  a hop size of $512$. The number of channels of the convolution layers in the encoder are $64, 128, 128, 256, 256, 256$, respectively. The number of channels of the decoder are the same as the encoder in reverse order. All (de-)convolution of the encoder (decoder) use $3 \times 3$ kernel size, with stride $2 \times 2$, and no padding. Encoder (decoder) layers are comprised of the (de-)convolution, with batch normalization, and a SELU activation function. For our GCN we use two layers with number of hidden units to be the same as the dimension of the embedding space.

\begin{table*}[!t]
    \centering
    \resizebox{1.98\columnwidth}{!}{
    \begin{tabular}{l?r|r|r?r|r|r?r|r|r?r|r|r?r|r|r}\toprule\midrule
        \multirow{3}{*}{Method} & \multicolumn{3}{c?}{Input SNR}  & \multicolumn{3}{c?}{Input SNR} & \multicolumn{3}{c?}{Input SNR} & \multicolumn{3}{c?}{Input SNR} &  \multicolumn{3}{c}{Input SNR}\\
         & \multicolumn{3}{c?}{-7.5 dB}  & \multicolumn{3}{c?}{-5.0 dB} & \multicolumn{3}{c?}{0 dB} & \multicolumn{3}{c?}{5.0 dB} &  \multicolumn{3}{c}{7.5 dB} \\
        \cmidrule(l{0.5em}){2-16}
                             & STOI & PESQ & SDR & STOI & PESQ & SDR & STOI & PESQ & SDR & STOI & PESQ & SDR & STOI & PESQ & SDR\\
        \midrule\midrule 
        Noisy               & 0.51 & 1.22 & -6.05 & 0.58 & 1.51 & -3.34 & 0.65 & 1.74 & 0.30 & 0.71 & 1.91 & 5.03 & 0.78 & 2.09 & 6.96 \\
        CRNN-C~\cite{chakrabarty2019time}  & 0.55 & 1.28 & -1.08 & 0.62 & 1.57 & 0.82 & 0.68 & 1.79 & 2.61 & 0.73 & 1.97 & 6.36 & 0.80 & 2.13 & 7.32 \\
        \midrule[1pt]
        Proposed & 0.60 & 1.67 & 3.84 & 0.66 & 1.98 & 5.74 & 0.72 & 2.11 & 6.84 & 0.77 & 2.24 & 8.86 & 0.81 & 2.36 & 9.94 \\
    \end{tabular}}
    \caption{Results (STOI, PESQ and SDR) for different SNR levels using a 4-microphone linear array.}
    \label{snr_results}
\end{table*}

\begin{figure}
    \centering
        \includegraphics[width=0.67\linewidth]{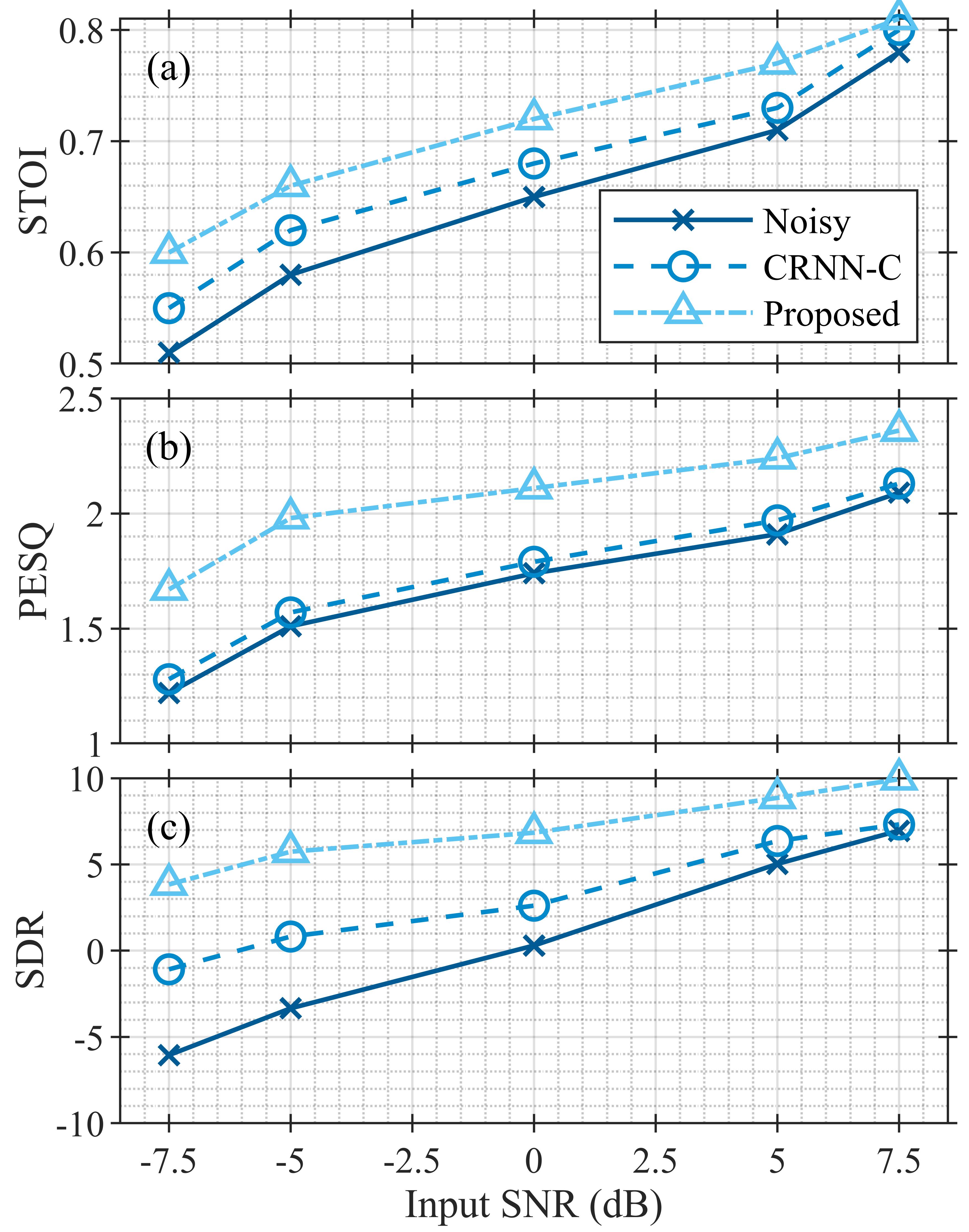} 
    \caption{STOI, PESQ, and SDR trends for different SNR levels.}
    \label{fig:SNRin_trend}
    \vspace{-0.2in}
\end{figure}

\subsection{Enhancement Results}

We compare our approach, with Chakrabarty's et al. (CRNN-C)~\cite{chakrabarty2019time}, a recently proposed multi-channel deep learning based enhancement model. Their approach uses as input the phase and magnitude of the raw signal for each channel and performs convolution across the channels to predict the ideal ratio mask (IRM) to compute the clean magnitude. We also show results of the U-Net architecture without utilizing the spatial information, i.\,e., using a single-channel for the enhancement. Table \ref{sim_results} shows results for different array configurations. 
 
Our method outperforms, the single-channel approach with high margins on all metrics. In addition, our model also surpasses the CRNN-C method. 
Surprisingly, we observe that CRNN-C approach is the worst performing, with values marginally above the noisy signal, even when comparing with the single channel method. 
In addition, our approach is robust against different microphone geometries in contrast with conventional methods where human a-priori knowledge is essential to achieve high-performing models.

Finally, in Table~\ref{snr_results} and Fig.~\ref{fig:SNRin_trend}, we analyze results at different SNR levels. Results are shown for a linear array with $4$ microphones at 5 different SNR levels, $[-7.5, -5,$ $0, 5, 7.5]$\,(dB). In general higher performance gains are obtained for negative SNR values compared to the positive ones. For -7.5\,dB input SNR, we see that the proposed GCN-based method leads to more than 9 dB improvement in SDR over the noisy case. For the highest input SNR, we observe only 2.98 dB improvement in SDR. In addition, we observe that our model outperforms CRNN-C on all SNR values. This is again more clear in very low SNR values such as $-7.5$\,dB, where our model has $0.5$, $0.39$, $4.92$ improvement on STOI, PESQ, and SDR, respectively.

\begin{table}[!t]
    \centering
    \begin{tabular}{l|c|c |c}\toprule\midrule
         Method                  & PESQ & STOI & SDR  \\\midrule
         Noisy                   & 1.72 & 0.66 & 0.19 \\
         $\mathcal{L}_{Mag}$     & 2.03 & 0.70 & 6.22 \\
         $\mathcal{L}_{Spec}$    & 1.88 & 0.69 & 6.28 \\
         $\mathcal{L}_{Mag+Spec}$    & 2.02 & 0.70 & 7.40 \\
         $\mathcal{L}_{Mag+raw}$ & 2.13 & 0.73 & 7.73 \\
    \end{tabular}
    \caption{Results on the development set for different loss functions.}
    \label{loss_results}
    \vspace{-0.15in}
\end{table}
\begin{table}[!t]
    \centering
    \begin{tabular}{l|c|c |c}\toprule\midrule
         Method  & PESQ & STOI & SDR  \\ \midrule
         Noisy   & 1.72 & 0.66 & 0.19 \\
         w/o GCN & 2.08 & 0.71 & 7.05 \\
         w/ GCN  & 2.13 & 0.73 & 7.73 \\
    \end{tabular}
    \caption{Proposed model with and without the graph representation.}
    \label{ablation_gcn}
    \vspace{-0.15in}
\end{table}
\subsection{Ablation Study}
We perform an ablation study using a 4-microphone linear array to find an appropriate loss function, and to verify that the GCN improves the model's performance. 
Experiments in this section use an 8 microphone circular array and results are shown with STOI, PESQ, and SDR.
Table~\ref{loss_results} shows the results on the development set of our simulated dataset. The results indicate that the loss that combines the magnitude and the raw signal have the overall best performance. 

To verify  the utility of the GCN in the embedding space of the U-Net, we perform two experiments where we: (i) discard the GCN, and (ii) include GCN, from the embedding space of U-Net. Table~\ref{ablation_gcn} depicts the results. Including GCNs the performance of the model is better on all the metrics than without using GCN. We should note that when we do not use GCN we still use all channels as we combine the extracted U-Net representations with the attention layer.

\vspace{-0.1in}
\section{Conclusions}
\vspace{-0.1in}
In this paper, we propose to utilize graph neural networks to exploit the spatial correlations in the multi-channel speech enhancement problem. We use a U-Net type architecture where the encoder learns representations for each channel separately, and a graph is constructed using these representations. Graph convolution networks are used to propagate messages in the graph, and hence learn spatial features. The features of each node are passed to the decoder to reconstruct the spectrogram of each channel. We combine these using attention weights for the final prediction of a reference microphone.
An analysis of the proposed method is provided with different array geometries and microphone counts in reverberant and noisy environments. This is the first study that utilizes GCN for speech enhancement. Results show the superiority of the proposed approach compared to the prior state-of-the-art method.


Future work could look at performing a quantitative evaluation of the trained model by inspecting the most important nodes and edges in the graph. 
Also, performing speech enhancement by using the raw waveform (i.\,e., end-to-end approach) instead of the complex spectrograms is another possible direction.

\vfill\pagebreak


\bibliographystyle{IEEEbib}
\bibliography{refs}

\end{document}

%% file: math_commands.tex
\usepackage{amsmath,amsfonts,bm}

\def\eqref#1{equation~\ref{#1}}

\def\1{\bm{1}}

\DeclareMathAlphabet{\mathsfit}{\encodingdefault}{\sfdefault}{m}{sl}
\SetMathAlphabet{\mathsfit}{bold}{\encodingdefault}{\sfdefault}{bx}{n}

